\documentclass[12pt]{article}
\usepackage{amsmath,amsthm,amsfonts,amssymb,amscd}
\usepackage{graphics}
\usepackage[latin1]{inputenc}

\begin{document}

\newcommand{\la}{{\lambda}}
\newcommand{\ro}{{\rho}}
\newcommand{\po}{{\partial}}
\newcommand{\ov}{\overline}
\newcommand{\re}{{\mathbb{R}}}
\newcommand{\nb}{{\mathbb{N}}}
\newcommand{\Z}{{\mathbb{Z}}}
\newcommand{\Uc}{{\mathcal U}}
\newcommand{\gc}{{\mathcal G}}
\newcommand{\hc}{{\mathcal M}}
\newcommand{\fc}{{\mathcal F}}
\newcommand{\dc}{{\mathcal D}}
\newcommand{\al}{{\alpha}}
\newcommand{\vr}{{\varphi}}
\newcommand{\om}{{\omega}}
\newcommand{\La}{{\Lambda}}
\newcommand{\be}{{\beta}}
\newcommand{\te}{{\theta}}
\newcommand{\Om}{{\Omega}}
\newcommand{\ve}{{\varepsilon}}
\newcommand{\ga}{{\gamma}}
\newcommand{\Ga}{{\Gamma}}
\newcommand{\zb}{{\mathbb{Z}}}
\def\sen{\operatorname{sen}}
\def\Ker{\operatorname{{Ker}}}
\newcommand{\bc}{{\mathcal B}}
\newcommand{\lc}{{\mathcal L}}
\newcommand{\ec}{{\mathcal E}}
\newcommand{\oc}{{\mathcal O}}

\centerline{\Large\bf Studies on Polyakov and Nambu-Goto Random Surface}

\vskip .1in 

 \centerline{\Large\bf  Path Integrals on Q.C.D. $(SU(\infty))$:
 Interquark Potential}
 
 \vskip .1in

 \centerline{\Large\bf  and Phenomenological Scattering Amplitudes}

\vskip .5in

\centerline{{\sc Luiz C.L. Botelho}}

\centerline{Departamento de Matemática Aplicada,}

\centerline{Instituto de Matemática, Universidade Federal Fluminense,}

\centerline{Rua Mario Santos Braga}

\centerline{24220-140, Niterói, Rio de Janeiro, Brazil}

\centerline{e-mail: botelho.luiz@ig.com.br}

\vskip .6in

\begin{abstract}
We present news path integral studies on the Polyakov Non-Critical and Nambu-Goto critical strings theories and its applications to Q.C.D. $(SU(\infty))$ interquark potential.

We also evaluate the long distance asymptotic behavior of the interquark potential on the Nambu-Goto string theory with an extrinsic term and in Polyakov's string at $D\to\infty$. We also propose an alternative and new view to covariant Polyakov's string path integral with a fourth-order two-dimensional quantum gravity, useful as an effective stringy  description for Q.C.D.$(SU(\infty))$ at the deep infrared region.
\end{abstract}

\section{Introduction}

One of the most promising mathematical formalism for a physically sensible description of strong interactions is quantum chromodynamics. In strong interaction physics the image of an physically detectable mesonic quantum excitations is, for instance, the quantum mechanical color invariant probability of the appearance of a pair quark-antiquark bounding a space-time non abelian gluon surface connecting both the pair's particle.

It appears tantalizing for mathematical formulations to consider as fundamental gauge invariant dynamical variable, the famous quantum Wilson loop, with the loop $C$ (defining the non-abelian holonomy factor); being given by the quark-antiquark (space-time) Feynman trajectory ([1], [2]).

It is thus searched loop space dynamical equations (at least on the Lattice on the formal grounds) for the quantum Wilson loop wich supports hopes for a complete bosonic random surface-bosonic string solution for euclidean Q.C.D., at least at the large number of colors ([1]).

It is natural thus to study path integral geometrodynamical random surfaces-string propagators in order to make such connection between string and QCD mathematically more precise ([3]), at least at the infrared region.

This paper is organized as follows: In section 2, we present a detailed stuty of the Nambu-Goto string path integral. In section 3 and appendixes A and B, we extend the results of section 2 to the case of the presence of extrinsic geometry. In the section 3 also, we present a new result on the large distance asymptotic behavior of the interquark potential by supposing QCD represented by the Nambu-Goto extrinsic string at large distance.

In section 4 and appendix C, we present a new proposal for the Polyakov's Non-Critical String.

In section 5, we deduce the interquark potential in the Polyakov's covariant string path integral at the limit $D\to+\infty$.

In section 6, we propose surface (string) area functionals as non perturbative quantum states of the $QCD(SU(\infty))$ vacuum structure.

\section{Basics Results on the Classical Bosonic Surface Theory and the Nambu-Goto String Path Integral}

Let us start our considerations by considering a given continuously differentiable globally orientable; compact surface $S$, with a boundary given by a non self-intersecting smooth curve $C$ and fully immerse on the space-time $R^D$.  Its mathematical description is described by a $C^1(\Om)$ two-dimensional vector field $X_\mu(\xi_1,\xi_2)$, $(\mu=1,\dots,D)$ on a two-dimensional domain $\Om$ (compatible with the fixed-prescribed topology of $S$).

The Nambu-Goto-Buff-Lovett-Stillinger ([1]) area is given by the usual geometric integral $((\xi_1,\xi_2) := \xi)$
\begin{equation}
A(S_c)=\int_\Om (\det\{h_{ab}(\xi)\})^{1/2} d^2\xi\tag{1}
\end{equation}
with $h_{ab}(\xi)=(\po_a X^\mu\po_b X_\mu)(\xi)$ denotes the metric tensor induced on the surface $S_c$. 

Note that the scalar of curvature 
$R(X^\mu)=\{h^{ab}(Ric)_{ab}\}(X^\mu(\xi)); \Big[ Ric_{ab}(X^\mu(\xi))= R_{abcd} = \frac12 \Big( \frac{\po^2 h_{ac}}{\po x^b \po x^d} - \frac{\po h_{bc}}{\po x^a \po x^d} - \frac{\po^2}{\po x^b \po x^c}+ \frac{\po^2 g_{cd}}{\po x^a \po x^c} + g_{rs} (\Ga_{ac}^r \Ga_{bd}^s - \Ga_{ad}^r \Ga_{bc}^s) \Big)$ $R_{acb}^c(X^\mu(\xi))=(\po_c\Ga_{ba}^c-\po_b\Ga_{ca}^c+\Ga_{ba}^d\Ga_{cd}^c-\Ga_{ca}^d\Ga_{bd}^c) (X^\mu(\xi))$, with naturally $\Ga_{bc}^a=(\frac12 h^{ar}(\frac\po{\po \xi^b} h_{rc}+\frac\po{\po\xi^c} h_{rb}-\frac\po{\po\xi^r} h_{bc})) \Big]$; must satisfy the topological constraint that originates from the Gauss theorem. For the case of boundaryless compact surfaces without ``handles''
\begin{equation}
\frac1{2\pi}\left( \int_\Om(\sqrt h\, R)(X^\mu(\xi)) d^2 \xi \right) = 2-2g, \tag{2}
\end{equation}
where $g$ denotes the number of ``holes''\, of $S_c$.

Note again that the domain $\Om$ must be compatible with the surface topology ([4]).

The most important property of the above written functionals on the Riemman surface $S_c$ \{(we point out that under the above mathematical imposed conditions on the surface $S_c$ added with an additional assumption of $C^\infty$-differentiability, it is a mathematical consequence that $S_c$ may be endowed now with a complex structure, turning it a Riemman surface, paramtrized by holomorphic vector fields on $\Om\subset R^2$; $X^\mu(\xi_1,\xi_2)=X^\mu(z)$ with $z=\xi_1+i\xi_2$, $i=\sqrt{-1}$)\} is the local invariance under the group of local diffeomorphism of the surface $S$
\begin{equation}
\begin{aligned}
\xi_a' -\xi_a &=\delta\xi_a=: \ec_a(\xi) \\
\delta X_\mu(\xi) &= \ec_a(\xi)\po^a X_\mu(\xi). \end{aligned} \tag{3}
\end{equation}

It is worth that the above pointed out invariance under the diffeomorphism local group of the surface $S_c$, can be extended to the global diffeomorphism case, only for trivial topological surfaces with $g=0$ (no ``handles'') (see eq.(2)).

Let us thus follow R.P. Feynman in his theory of path integration sum over ``classical-random''\, histories of a quantum system (with a classical mechanical system counter-part as in our case: The Correspondence  Principle of Quantum Mechanics in action), in order to quantize our string theory (the curve $C$ can be considered as our classical string and the surface $S_c$ denotes its euclidean quantum trajectory in the space-time $R^D$), through an anihillation string process).
\begin{align*}
G(C) &= \sum_{g=0}^\infty \left\{\int_{X_\mu(\xi)|_{\xi\in\po\Om}=C^\mu(\sigma)} d_h\mu[X_\mu(\xi)] \times \exp \left\{-\frac1{2\pi\al'} A(S_c)\right\} \right. \\
&\quad \left. \times \delta^{(F)} \left( \left[ \int_C K_h(\sigma)d\sigma+\frac1{2\pi} \int (\sqrt h R)(X^\mu(\xi))-(2-2g)\right] \right) \right\}. \tag{4}
\end{align*}

Here $\al'$ denotes the Regge slope parameter which has the dimension of inverse of mass square (in universal units $\hbar=c=1$). Here $K_h(\sigma)$ is the geodesic curvature of the surfaces boundary $C$.

The above written non-polinomial $2D$ quantum field theory is somewhat complicated in its methematical perturbative calculational structure, since $d\mu_h(X^\mu(\xi))$ is not the usual Feynman product measure, but the ``weighted''\, product Feynman measure, in order to preserve the invariance of the geometrodynamical string propagator eq.(4) under the action of the main theory's symmetry eq.(3). Explicitly
\begin{equation}
d_h\mu[X^\mu(\xi)]:= \prod_{\xi\in\Om}[\big(h(X(\xi))\big)^{1/4} \, dX_\mu(\xi)].\tag{5}
\end{equation}

The above written local diffeomorphism invariant functional measure on the string vector position is obtained form the local diffeomorphism invariant functional Riemann metric
\begin{equation}
||\delta X_\mu||^2=\int_\Om(h(X^\mu(\xi))^{1/2}(\delta X_\mu)(\xi)(\delta X_\mu)(\xi) d^2 \xi . \tag{6}
\end{equation}

However the first step to evaluate the so called Nambu-Goto string path integral eq.(4) is to consider the quantization process as ``quantum fluctuations''\, around the classical system motion (R.P. Feynman):
\begin{equation}
X_\mu(\xi)=X_\mu^{CL}(\xi)+\left( \sqrt{\pi \al'} \right) \ov X_\mu(\xi)\quad (\hbar=1).\tag{7}
\end{equation}

The classical dynamics is given by Euler-Lagrange equations associated to the surface area functional under the topological constraint eq.(2). And before proceeding, we remark that is only in this step and on the topological form of $\Om$ where the topological constraint is taken into account. As a consequence we can disregard explicitly the functional topological constraint on eq.(4). We have thus, the classical motion equations (a Dirichlet nonlinear problem)
\begin{equation}
\begin{aligned}
\, & \Delta_h X_\mu^{CL}(\xi)=0 \\
& X_\mu^{CL}(\xi)|_{\xi\in\po\Om}=C^\mu . \end{aligned} \tag{8}
\end{equation}

Here $\Delta_h$ denotes the second-order elliptic operator called Laplace-Beltrami associated to the metric 
$$
h_{ab}(\xi)=h_{ab}(X_\mu^{CL}(\xi))=(\po_ a X^{\mu,CL} \po_b X_\mu^{CL})(\xi).
$$
\begin{equation}
\Delta_h = \left( \frac1{\sqrt h} \po_a(\sqrt h \, h^{ab} \po_b) \right)(\xi).\tag{9}
\end{equation}

At this point, one most use formally, at least for trivial topological surfaces $S_c$, the hypothesis of the global extension of the local diffeomorphism group of the surface $S_c$, in order to use globally on $S_c$ the conformal coordinate system 
$$
\xi_1'=\xi_1'(\xi_1,\xi_2);\quad \xi_2'=\xi_2'(\xi_1,\xi_2).
$$
\begin{equation}
h_{ab}(X^{'\mu}(\xi'))=\rho^2(X^{'\mu}(\xi'))\delta_{ab}.\tag{10}
\end{equation}

The new coordinates are given by the global (or local) diffeomorphic solutions of the Laplace-Beltrami: equations given below $(i=\sqrt{-1})$
\begin{align*}
d\xi_1' +id\xi_2' &= \la(\xi_1,\xi_2) \Big( \sqrt{h_{11}(X_\mu(\xi))}\, d\xi_1 \\
&\quad+ \frac{(h_{12}(X_\mu(\xi))+i\sqrt{h(X_\mu(\xi))})}{\sqrt{h_{11}(X^\mu(\xi))}}\, d\xi_2 \Big) \tag{11-a}
\end{align*}
\begin{align*}
d\xi_1'-id\xi_2' &= \mu(\xi_1,\xi_2) \Big( \sqrt{h_{11}(X_\mu(\xi))}\, d\xi_1 \\
&\quad+ \Big( \frac{(h_{12}(X_\mu(\xi))-i\sqrt{h(X_\mu(\xi))})}{\sqrt{h_{11}(X^\mu(\xi))}} \Big) d\xi_2 \, \Big).
\tag{11-b}
\end{align*}

In this new coordinate system on $S_c$ ([4])
\begin{equation}
h_{ab}(X_\mu'(\xi'))=\rho^2(\xi')\delta_{ab}=\left(\frac1{|\la\mu|(\xi(\xi'))} \right) \delta_{ab}. \tag{12}
\end{equation}

If one now choose the conformal gauge for the surface $S_c$ \, $h_{ab}(\xi)=e^{\vr(\xi)} \delta_{ab} (\vr(\xi)=2 \ln \rho(\xi))$, one reduces the string non Linear elliptic problem eq.(8) to the well-studied Dirichlet problem in $\Om$
\begin{equation}
\begin{cases}
\Delta_{h=\delta_{ab}} X_\mu'(\xi')=0 \\
X_\mu'(\xi')|_{\xi'\in\po\Om'} = C^{'\mu} . \end{cases}\tag{13}
\end{equation}

Note that the problem full solution is given by
\begin{equation}
X_\mu^{CL}(\xi)=X_\mu'(\xi'(\xi)).\tag{14}
\end{equation}

The solution of eq.(13) can be always be analyzed by methods of conformal complex variable methods ([4], [5]) specially for the trivial topological case (connected planar $\Om'$) ([4]).

Unfortunatelly, it appears that the resulting quantum theory (path integral) for eq.(4)--eq.(5) still remains as an open problem, from a purely perturbative approach around the loop-expansions in a $\al'$ -- power expansion.\footnote{$$
a)\qquad h_{ab}(\xi) = \left( \overbrace{\po_a X^{\mu,CL} \po_b X^{\mu,CL}}^{h_{ab}^{CL}} + (\pi \al') \overbrace{\po_a \overline X^\mu \po_b \overline X^\mu}^{h_{ab}^q} \right) (\xi)
$$
$$
\qquad\qquad\quad b)\qquad \sqrt{h(\xi)} = \sqrt{h^{CL}(\xi)} \exp \left\{ - \frac12 \sum_{n=1}^\infty \frac{(-1)^n (\pi \al')^n}{n} (h_{ab}^{-1,CL} \cdot h_{ba})^n(\xi) \right\}
$$
}

However in the simply case of the domain $\Om_{(R,T)}$ being a rectangle of sizes $0\le\xi_1\le R$; $0\le\xi_2\le T$, an exact very useful one-loop result can be obtained (Note that $C=\po\Om$).

It is straightforward to see that
\begin{align*}
S[\ov X^\mu(\xi_1,\xi_2)] &= \frac1{2\pi\al'}(RT) + \\
&\quad+ \frac12\int_0^T d\xi_2 \int_0^R d\xi_1 (\po \ov X^\mu(\xi))^2+O(\al^{'2}).\tag{15-a}
\end{align*}
\begin{equation}
d_h\mu[\ov X^\mu(\xi_1,\xi_2)]=D^F[\ov X^\mu(\xi_1,\xi_2)]+O(\al'). \tag{15-b}
\end{equation}

This leads to the result
\begin{align*}
G(C_{[R,T]}) &= \exp \left\{ - \frac1{2\pi\al'} RT \right\} \\
&\quad \times \left( \det_{\Om_{(R,T)}}(-\Delta) \right)^{-(\frac{(D-2)}2)}. \tag{16}
\end{align*}

By noting the explicit (non-trivial) evaluation of the functional determinant of the Laplacean on the Torus $\Om_{(R,T)}$ with Dirichlet conditions ([7], vol II):
\begin{align*}
\, & \left(\det_{\Om_{(R,T)}}(-\Delta) \right)^{-1/2} \\
&\quad = \frac1{(\frac TR)^{1/2}} \Big[ \left( e^{-\frac{4\pi T}R}\right)^{-1/24} \\
&\qquad \times \left( \prod_{n=1}^\infty(1-e^{-\frac{2\pi nT}R})^{-2} \right) \Big], \tag{17}
\end{align*}
one can see that the quantum strong ground state has the ``confining behavior''\, with the Coulomb-Lüscher term as its energy on this one-loop approximation
\begin{equation}
E_{\text{Vacuum}}(R)=\lim_{T\to+\infty} \left\{ - \frac1T \ln G(C_{(R,T)})\right\} = \left(\frac R{2\pi\al'}\right)-\left(\frac{\pi(D-2)}6 \cdot \frac1R \right).\tag{18}
\end{equation}

Unfortunatelly ths string scattering amplitudes were never evaluated in a undisputable form in this Nambu-Goto string theory, unless on the light-cone gauge by S. Mandelstam ([3], [6]).

As a consequence of the above mathematical aspects on the Nambu-Goto quantum string theory, A.M. Polyakov has proposed a new functional integral approach to overcame some of the above difficulties.

The complete mathematical exposition of the A.M. Polyakov propose will be exposed (in details) on next section.

As a final comment let us try to evaluate the Nambu-Goto string path integral through the use of the conformal gauge as given by eq.(12). In this case (the so called light -- cone gauge), we have the following constraints, after introducing the complex-light-cone euclidean coordinate on the domain $\Om$
\begin{equation}
\begin{aligned}
\po_+ &= \frac{(\po_{\xi_1'}-i\po_{\xi_2'})}2 = \po_z \\
\po_- &= \frac{(\po_{\xi_1'}+i\po_{\xi_2'})}2 = \po_{\bar z} . \end{aligned} \tag{19} 
\end{equation}

We have that 
\begin{equation}
\po_+X_\mu'\po_+X_\mu'=\po_-X_\mu'\po_-X_\mu'(\Leftrightarrow \po_{\xi_1'} X_\mu'\po_{\xi_2'}X_\mu'=h_{12}=h_{21}=0) \tag{20-a}
\end{equation}
\begin{equation}
\sqrt{h(X_\mu'(\xi'))} = \po_+ X_\mu'\po_-X_\mu'(\Leftrightarrow h_{11}=h_{22}=\po_+X_\mu'\po_-X_\mu').\tag{20-b}
\end{equation}

In this gauge the path integral eq.(4) for $C=\{\phi\}$ (the string partition functional) takes the form (for the simple case of trivial topology surface $g=0$).
\begin{align*}
Z &= \int d_h\mu[X_\mu'(z)] \exp \Big\{ - \frac1{2\pi\al'} \int_\Om \left( \frac{dz\wedge d\bar z}{2i} \right) \\
&\quad \times (\po_z X^{'\mu})(\po_{\bar z} X^{'\mu})(z,\bar z) \Big \} \\
&\quad \times \delta^{(F)} [(\po_z X_\mu' \po_z X_\mu')] \times \delta^{(F)} [(\po_{\bar z} X_\mu' \po_{\bar z} X_\mu')]. \tag{21}
\end{align*}

Note that in the practical use of eq.(1) to evaluate string observable average, one already uses the observable on the light-cone gauge eq.(20), which by its turn suppress the explicitly use of the above written delta functionals insuring that Feynman-Wiener measure $d_h\mu[X_\mu'(\xi)]$ is already in this gauge.

It has been proved by my self ([1]) that the non-linear measure $d_h\mu[X_\mu'(\xi)]$ on the light-cone gauge can be related by the A.M. Polyakov-Strominger-Fujikawa conformal anomaly factor to the Feynman-Wiener simply weighted measure
\begin{align*}
\, & d_h\mu[X_\mu'(\xi)]= \left( \prod_{(\mu,z,\bar z)} dX_\mu(z,z')\right) \times \\
&\quad \times \left\{ \frac{\delta^{(2)}(0)}4 \int_\Om\ln(h(X_\mu(z,z)))dz\, d\bar z\right\} \times \\
&\quad \times \exp\left\{ - \frac{(26-D)}{48\pi} \int_\Om \frac{dzd\bar z}2 \left[ \frac{\po_+(\po_+X_\mu\po_-X_\mu) \cdot \po_-(\po_+X_\mu\po_-X_\mu)}{(\po_+ X_\mu\po_- X_\mu)^2} \right](z,z) \right\}. \tag{22}
\end{align*}

One can see thus that only at $D=26$, the pure bosonic Nambu-Goto string is described by massless scalar fields on the string domain $\Om$,
if one uses formally the Bollini-Giambiagi dimensional regularization scheme to assign the unity value for the tad-poles Feynman diagramms\footnote{\,\,\,If $\Delta(\xi)=\int \frac{d^Dk}{(2\pi)^{D/2}}\, \frac{e^{ik\xi}}{k^\be}$ with $\be\in R$, then on the dimensional regularization scheme $\Delta(0)=\int d^D k|K|^{-\be}=0$, even if $\be=0$.}
([1], [4]),
through a not completely understood Feynman diagrammatics for two-dimensional (mathematically ill defined) Massless scalar fields for $\Om=R^2$ (the so called Coleman theorem ([1])).

However, One can follows the Virassoro-Sakita proposal to evaluate string scattering amplitudes using scalar vertex without bothering ourselve with gauge fixed technical details ([3]).

We thus use the path integral eq.(21) for $\Om=R^2$ in order to evaluate the closed string (scalar) $N$-point scattering amplitude at $D=26$ (with $\delta^{(2)}(0)=0$) 
\begin{align*}
A(\rho_1^\mu,\dots,\rho_N^\mu) &= \frac1Z \Big\{ \int D^F[X_\mu(\xi)] \exp \left[-\frac1{2\pi\al'} \int_{R^2} (\po X_\mu\cdot\po X_\mu)(\xi) d^2\xi \right] \\
&\quad \times \left[ \int_{R^{2N}} d^2 \xi_j \prod_{j=1}^N \exp(i\rho_\mu^j\cdot X_\mu)(\xi_j) \right] \Big\}. \tag{23}
\end{align*}

Since eq.(23) is formally a Gaussian Functional Integer and we can re-write the scalar vertexs as string vector position source $(i=\sqrt{-1})$
\begin{align*}
\, & \left( \prod_{j=1}^N \exp(i\rho_\mu^j X_\mu)(\xi_j)\right) \\
&= \exp \left\{ i \left[\sum_{j=1}^N \int_{\re^2}(\rho_\mu^j \delta^{(2)}(\xi-\xi_j))X_\mu(\xi) d^2\xi \right] \right\}, \tag{24}
\end{align*}
and by using the dimensional  regularization technique to vanish the tad-pole term
\begin{equation}
\exp\left\{-\sum_{j=1}^N(\rho_\mu^j)^2(-\Delta)^{-1}(\xi_j,\xi_j) \right\}=1.\tag{25}
\end{equation}

One gets the well-known closed Veneziano $N$-point amplitude as a result in $R^{26}$ ([3]. [6])
\begin{align*}
A(\rho_1^\mu,\dots,\rho_N^\mu) &= \int_{R^{2N}} d^2 \xi_1\dots d^2 \xi_N \times \\
&\quad \times \prod_{i<j}^N \left( |\xi_i-\xi_j|^{\frac{(\rho_i^\mu, \rho_j^\mu)}{\pi\al'}} \right) \tag{26}
\end{align*}

\section{The Nambu-Goto Extrinsic Path String}

Another important classical local diffeomorphism invariant surface functional, closely related to the Q.C.D. $(SU(\infty))$ string ([1]) is the so-called extrinsic functional which is defined by the square of the surface mean curvature (see eq.(9)), with $\ga\in R$
\begin{equation}
\fc_{\text{exta}}(S)=\ga^2\int_\Om[\sqrt h\, (\Delta_h X_\mu)^2 ](\xi) d^2\xi.\tag{27}
\end{equation}

Let us add such functional to the area functional eq.(1) and consider the associated string path integral propagator eq.(4) (in the trivial topological sector of surface $S_c$).
\begin{align*}
\ov G(C) &= \int_{X_\mu(\xi)|_{\xi\in\po\Om}=C^\mu(\sigma)} d_h\mu[X_\mu(\xi)] \\
&\quad \times \exp\left\{ -\frac1{2\pi\al'}\int_\Om(h(X_\mu(\xi)))^{1/2} d^2\xi \right\} \\
&\quad \times \exp \left\{-\ga^2\int_\Om[\sqrt h(\Delta_hX_\mu)^2](\xi) d^2\xi \right\}. \tag{28}
\end{align*}

Let us note that even by using the light-cone string coordinate system, One still has a non-polinomial interacting quantum two-dimensional $SO(D)$ scalar field theory.

Namelly
\begin{align*}
\ov G(C) &= \int d_h\mu[X_\mu'(z,\bar z)] \times \\
&\quad \times \exp\left\{-\frac1{\pi\al'} \int_\Om dzd\bar z\left( \frac12(\po_+ X^{'\mu})(\po_- X^{'\mu})\right)(z,\bar z) \right\} \times \\
&\quad \times \exp\left\{ -\ga^2\int_\Om dzd\bar z \left[ \frac{(\po_+^2 X^{'\mu})(\po_-^2 X{'\mu})}{(\po_+ X^{'\mu})(\po_- X^{'\mu})} \right] \right\} \times  \\
&\quad \times \delta^{(F)} [(\po_+ X^{'\mu} \cdot \po_+ X^{'\mu})] \times \\
&\quad \times \delta^{(F)} [(\po_- X^{'\mu} \cdot \po_- X^{'\mu})]. \tag{29}
\end{align*}

However one can evaluate formally scattering amplitudes as done previously (see eq.(23)) by considering the string quantum trajectory as a small perturbation around the flat metric. The outcome is the following one-loop string scalar scattering amplitude that should be compared with the Nambu-Goto case of the previous section eq.(23)
\begin{align*}
A(\rho_1^\mu,\dots,\rho_N^\mu) &= \frac1Z \Big\{ \int D^F[X_\mu(\xi)] \times \\
&\quad \times \exp\left[-\frac1{\pi\al'}\int_{R^2}\frac12(\po X_\mu\po X_\mu)(\xi)d^2\xi \right] \times \\
&\quad\times \exp\left[-\ga^2\int_{R^2}(\po^2 X_\mu)(\po^2 X_\mu)(\xi) d^2\xi \right] \times \\
&\quad\times \left[\int_{R^{2N}} d^2\xi_j\left(\prod_{j=1}^N \exp(i \rho_\mu^j X_\mu)(\xi_j)\right) \right] \Big\}. \tag{30}
\end{align*}

Since eq.(30) is a Gaussian Functional Integral, One obtains an exactly result, where we have used the result
\begin{align*}
\, & \left[ \frac1{(\pi\al')}(-\po^2)+\ga(\po^2)^2\right]^{-1}(\xi,\xi') \\
&\quad := \pi\al' \left[\left(-\frac1{2\pi}\ln|\xi-\xi'|\right)-\frac1{2\pi} \left( K_0 \left(\left( \frac1{\pi\al'\ga^2} \right)^{1/2} |\xi-\xi'| \right) \right) \right]. \tag{31}
\end{align*}

Namelly:
\begin{align*}
\, & A(\rho_1^\mu,\dots,\rho_N^\mu) \\
&= \int_{R^{2N}} d^2\xi_1 \dots d^2\xi N \\
&\quad \times \left(\prod_{i<j}^N(\xi_i-\xi_j)^{\frac{(\rho_i^\mu,\rho_j^\mu)}{\pi\al'}} \right) \\
&\quad\times \left( \exp \left[ - \frac{\al'}2 \sum_{i<j}^N K_0 \left(\left(\frac1{\pi\al' \ga^2}\right)^{1/2} |\xi_i-\xi_j|\right) \right]  \right). \tag{32}
\end{align*}

Unfortunatelly, it appears that the corrections coming from the string extrinsic functions do not modify the usual Veneziano-Virassoro bosonic closed string spectrum. But a clear proof of this no go result is still missing.

At this point we analyze the vacuum string energy for the domain $\Om_{R,T)}$ being a rectangle of sizes $0\le\xi_1\le R$; $0\le\xi_2\le T$ through a one-loop approximation in the Regge slope constant $\al'$ (see eq.(15-a) of the previous section).

In this case we have the exact result
\begin{align*}
\ov G[C_{(R,T)}] &= \exp \left\{-\frac1{2\pi\al'} RT \right\} \times \\
&\quad\times \left[ \det_{\Om_{(R,T)}} \left( \ga^2(-\po^2) \left( \frac1{\ga^2}+(-\po^2)\right) \right) \right]^{-\frac{(D-2)}2} .\tag{33}
\end{align*}

The evaluation of the fourth-order elliptic operator (with Dirichlet conditions) can be accomplished through the result
\begin{align*}
\, & \left[\det_{\Om_{(R,T)}} \left( \ga^2(-\po^2) \left( \frac1{\ga^2}+(-\po^2) \right) \right) \right]^{-\frac{(D-2)}2} \\
&\quad = \left[ \left( \det_{\Om_{(R,T)}}^{-\frac{(D-2)}2} [(-\po^2)]\right) \right. \\
&\qquad \times \left. \left( \det_{\Om_{(R,T)}}^{-\frac{(D-2)}2} \left[\frac1{\ga^2}+(-\po^2)\right] \right) \right]. \tag{34}
\end{align*}

But functional determinants have been evaluated in the literature ([7]). (See eq.([7])). The second order massive operator has the following result
\begin{align*}
\, & \det^{-\frac{(D-2)}2} \left[(-\po^2)+\frac1{\ga^2} \right] \\
&= \exp \Big\{ \frac{(D-2)\pi T}R \Big( \frac16-\frac R{\ga^22\pi} \\
&\quad+ \left(\frac R{2\pi\ga^2}\right)^2 \ln \left( \frac{4\pi e^{-\hat\ga}}R \right) \\
&\quad+ \frac12 \left( \frac R{2\pi\ga^2} \right)^4 \left( \int_0^1 dx(1-x)\sum_{n=1}^\infty(n^2+x \left( \frac R{2\pi\ga^2} \right)^2 \right)^{-\frac32}  \Big) \Big\} \\
&\quad\times \left\{ \prod_{n=-\infty}^{+\infty} \left[ \left[ 1-\exp \left(\left(-\frac{2\pi T}R \sqrt{n^2+\left( \frac R{2\pi\ga^2}\right)^2} \right) \right) \right] \right] \right\}^{-(D-2)}. \tag{35}
\end{align*}

Let us remark that $\left(t=\frac R{2\pi\ga^2}\right)$ (appendix B)
\begin{align*}
\, & \int_0^1 dx(1-x) \left[ \frac1{\left(n^2+x\left( \frac R{2\pi\ga^2} \right)^2 \right)^{3/2}} \right] \\
&= -\frac2{t^2} \left( \frac1{(n^2+t^2)^{1/2}} \right) + \frac{2n^2}{t^4(n^2+t^2)^{1/2}} \\
&\quad + \frac2{t^2n} - \frac2{t^4} (n^2+t^2)^{1/2}. \tag{36}
\end{align*}
which unfortunatelly leads to an -- in principle -- divergence on the string path integral $\ov G(C_{R,T})$ for the extrinsic string. However it is possible to get a somewhat phenomenological finite result for the vacuum energy (the interquark potential) for large $R$ $(R\to\infty)$
\begin{equation}
E_{\text{Vac}}(R)=\lim_{T\to\infty}\left\{-\frac1T \ln G(C_{(R,T)})\right\}.\tag{37}
\end{equation}

The point is that for large $R$ $(R\to\infty)$, the Epstein sum below written has a finite asymptotic behavior (see appendix A for a detailed evaluation), $s\in R$
\begin{align*}
\, & S_{\text{epst}}(s,a^2) := \sum_{n=1}^\infty \left( \frac1{(n^2+a^2)^s}\right) \\
&\quad = \frac1{\Ga(s)} \left( \int_0^\infty dU\, U^{s-1} \cdot 
e^{-U a^2} \left[ Tr_{\left\{
\begin{matrix}
C^2([0,1]) \\
f(0)=f(1)=0 \end{matrix} \right\}  }
 \left( e^{U(-\frac{d^2}{dz^2})} \right) \right] \right) 
\\
&\quad \overset{a\to\infty}{\sim} \left(\frac1{4\pi \Ga(s)} \right) \left[(a^2)^{-(S-\frac32+1)} \right]. \tag{38}
\end{align*}

As a consequence of the finite behavior eq.(38), One has the asymptotic (distributional L. Schwartz sense)
\begin{equation}
\sum_{n=1}^\infty \frac1{ \left(n^2+  x \left(\frac R{2\pi\ga^2} \right)^2 \right)} \overset{R\to\infty}{\sim}
\,\, \frac1{2(\pi)^{3/2} x\left( \frac R{2\pi\ga^2}\right)^2}\, . \tag{39}
\end{equation}

After substituting this (formal) result on eq.(35), by taking into account the (regularized) integration:
\begin{align*}
\, & \lim_{R\to\infty} \left\{\frac12\left(\frac R{2\pi\ga^2}\right)^4 \left[\int_0^1 dx(1-x)\left(\sum_{n=1}^\infty \frac1{(n^2+x \left( \frac R{2\pi\ga^2}\right)^2)^{3/2}} \right) \right] \right\} \\
&= \frac12\left( \frac R{2\pi\ga^2}\right)^4 \left[ \int_0^1 dx(1-x) \frac1{2\pi^{3/2} x \left(\frac R{2\pi\ga^2}\right)^2} \right] \\
&\sim \frac{\left(\frac R{2\pi\ga^2}\right)^2}{4\pi^{3/2}} \left( \int_{\ec_{QCD}}^1 \frac{dx(1-x)}x \right) \\
&= \frac{\left(\frac R{2\pi\ga^2}\right)^2}{4\pi^{3/2}} \left[ -\ell n(\ec_{QCD})-1 \right] \\
&= - \left\{ \left( \frac{(-\ell n(\ec_{QCD})+1)}{16\pi^{5/2} \ga^2} \right) \right\} R^2. \tag{40}
\end{align*}

Here $\ec_{QCD}$ is the underlying cut-off on the ``Feynman''\, parameters x $(\ec_{QCD}\le x\le 1)$. It is worth remark that the regularization parameters will be absorbed in the bare extrinsic coupling constant $\ga^2$ (somewhat related to the $Q.C.D(SU(\infty))$ coupling constant $\ga^2=(g_\infty)^2(\langle 0|F^2|0\rangle_{SU(\infty)})$. Namelly:
$$
\frac1{\ga_{\text{bare}}^2}(\ec)=\frac1{\ga_{\text{bare}}^2} ((\ln \ec_{QCD})+1).
$$

The contribution of this term to the phenomenological `´interquark potential''\, on the extrinsic string theory is exactly given below (to be added to the pure Nambu-Goto term eq.(18).
\begin{align*}
\, V^{\text{extrinsic}}(R) &= - \left( \frac{(D-2)}6 \cdot \frac\pi R\right)+\left( \frac{(D-2)}2 \cdot \frac1{\ga^2_{\text{ren}}} \right) \\
&\quad - \left(\left( \frac{(D-2)}{4\pi}\cdot \frac1{\ga_{\text{ren}}^4} \right) R\cdot \ln(4\pi e^{-\hat\ga}) \right) \\
&\quad+ \left( \frac{(D-2)}{4\pi} \left( \frac1{\ga_{\text{ren}}^4} \right) \,\, (R\ln R) \right) \\
&\quad + \left( \frac{(D-2)}{16\pi^{5/2}} \frac1{\ga_{\text{ren}}^2} \cdot R \right). \tag{41}
\end{align*}

We think that eq.(41) is an important result on applications of string representations for QCD, since it leads to a growing force that goes to infinite for $R\to\infty$, confining the static color quarks charge, certainly a real ``quark confinement'', opposite to the pure Nambu-Goto case with a ``weak confinement by a {\it constant\/} force. quarks are really confined if a string for QCD holds true at the limit of large $R$. So a QCD point particle description should only be phenomenological.

We  comment also that such kind of behavior for the interquark potential is compatible with a Mandelstan Gluonic propagator of the form ([8])
\begin{equation}
D_m(x-y)=\frac1{(2\pi)^{D/2}} \left( \int d^Dp \cdot e^{i\rho(x-y)} [(\ell n(|p|^2))/|p|^4] \right). \tag{42}
\end{equation}

\section{Studies on the perturbative evaluation of closed Scattering Amplitude in a Higher order Polyakov's Bosonic String Model}

It is well known that Polyakov covariant string theory with exactly soluble Liouville two dimensional model (the famous non critical string) has in its protocol the main hope of suppressing the tachionic excitation of the usual Nambu-Goto string ([10], [11], [12])\footnote{For a detailed presentation of the Polyakov's string path integral see \S3 of [9].}.

By the other side it is less known that Polyakov-Liouville effective string theory is somewhat phenomenological in the sense that it has already built in the assumption of ``weak''\, two dimensional induced quantum gravity, since one replaces its covariant path integral measure $D^F[e^{\vr/2}]\equiv\prod\limits_{\xi\in D} d(e^{\vr/2}(\xi))$ by the usual Feynman path measure $D^F[e^{\vr/2(\xi)}]\sim D^F[\vr(\xi)]$ when considering the final effective Liouville-Polyakov field theory on string world-sheet.

So in this approximate scheme of non critical strings (but quite useful ([2])), we in this section (somewhat pedagogical) introduce a somewhat toy model of higher-order two dimensional Polyakov covariant string with improved ultra violet behavior and show the exactly solubility of the resulting covariant Polyakov path integrations. 

Let us start our analysis by considering the theory's $N$-point off-shell closed Scattering amplitude defined by the following Polyakov's bosonic string general path integral
\begin{align*}
A(P_1^\mu,\dots,P_N^\mu) &= \int d^{\text{cov}} \mu[g_{ab}(\xi)] e^{-\frac{\mu^2}2 \int_{R^2}(\sqrt g)(\xi) d^2\xi} \\
&\quad \times \left[ e^{\overbrace{-\frac\ga2\int_{R^2}d^2\xi d^2\xi'\sqrt{g(\xi)} \sqrt{g(\xi')} R(\xi) K_g(\xi,\xi')R(\xi')}^{=\text{newproposed term}}} \right]
 \\
&\quad\times \left[\int d^{\text{cov}}\mu[X^\mu(\xi)]\exp\left\{ -\frac12\int_{R^2}(\sqrt g\, g^{ab} \po_a X^\mu \po_b X^\mu)(\xi)d^2\xi \right\} \right]
 \\
&\quad\times \left\{ \int_{R^{2N}}\prod_{i=1}^N \left[ \sqrt{g(\xi_i)} \cdot \exp(P_i^\mu X_\mu(\xi_i)) \right] d^2 \xi_i \right\}.\tag{43}
\end{align*}

Here the covariant functional measures are the functional volume elements associated to the functional Riemann metrics ([1], [2]. [3])
$$
||\delta g_{ab}||^2=\int_{R^2}\left[ \sqrt g ((\delta g_{ab})(g^{aa'} g^{bb'}+cg^{ab}g^{a'b'})(\delta g_{a'b'})) \right]  d^2\xi
$$
\begin{equation}
||\delta X^\mu||^2=\int_{R^2} [\sqrt g\, \delta X^\mu \delta X^\mu](\xi) d^2\xi. \tag{44}
\end{equation}

The Green function $K_g(\xi,\xi')$ of the Laplace-Beltrami operator on the presence of the metric $g_{ab}(\xi)=e^{\vr(\xi)} \delta_{ab}$, in the conformal gauge, with a covariant cut-off  already built in is given explicitly by the Riesz-Hadamand formula ([4], [5]) on $R^2$
\begin{equation}
K_\vr(\xi,\xi') = \begin{cases}
-\frac1{2\pi} \ln(|\xi-\xi'|) \qquad & \xi\ne\xi' \\
\frac1{2\pi}(\frac1\ec)-\frac1{2\pi} \vr(\xi)-\frac1{2\pi} \ln(\ec) & \xi=\xi' \end{cases} \tag{45}
\end{equation}

Note that eq.(3) is a direct result that in the conformal gauge $g_{ab}=e^{\vr(\xi)} \delta_{ab}$ the Laplace-Beltrami operator reduces to the usual Laplace operator for $\xi\ne\xi'$. And for those points $\xi=\xi'$, one should use the famous parameter formula of J. Hadamard for the Laplace Beltrami Operator on $\re^2$ togheter by taking into account the fact that the geodesic distance $S(\xi,\xi')$ in $R^2$, endowed with a metric in the conformal gauge is exactly given by $S(\xi,\xi')=e^{\vr(\xi)}|\xi-\xi'|$ for $\xi\to\xi'$.

In other words, for $\xi\to\xi'$, we have the regularized asymptotic behavior an $R^N$ ([5])
\begin{align*}
K(\xi,\xi') &\sim \lim_{N\to2} \left\{ \frac{\Ga(\frac12 N)}{2\pi^{(\frac N2)}(N-2)} e^{(2-N)\ln(S(\xi,\xi'))} \right\} \\
& \overset{{}_{\substack{N\to2\\ \ec\to0}}}{\sim} \frac1{2\pi} \left( \frac1{N-2}\right) - \frac1{2\pi} \ln \left( e^{\vr(\xi)}(|\xi-\xi'|^2+\ec^2)^{1/2} \right) \\
& \overset{|\xi-\xi'|\to0}{\sim} \frac1{2\pi\ec} - \frac{\vr(\xi)}{2\pi}-\frac1{2\pi}\ln(\ec).\tag{46}
\end{align*}

In order to evaluate perturbativelly eq.(1) around the flat metric $g_{ab}(\xi)\sim\delta_{ab}$, we consider the following approximation on the resulting effective functional measure on the metric field by replacing it by its weak metric feynman measure
\begin{align*}
d\mu[g_{ab}] &= D^F[e^{\frac\vr2(\xi)}] \\
&\quad \times \exp\left\{-\frac{26}{48}\int_{R_+^2} \frac12(\po\vr)^2(\xi)d^2\xi \right\} \\
&\quad\times\exp\left\{-\frac{\mu^2}2 \int_{R_+^2} e^{\vr(\xi)} d^2\xi \right\} \\
&\quad\times\exp\left\{-\frac{\ga^2}2\int_{R^2}(\po^2\vr)^2(\xi)d^2\xi\right\} \\
&\quad \overset{\text{(replace by)} }{\sim} D^F[\vr(\xi)]\exp\left\{ -\frac{26}{48}\int_{R^2} \frac12(\po\vr)^2d^2\xi\right\} \exp\left\{-\frac{\ga^2}2\int_{R^2}(\po^2\vr)^2(\xi) d^2\xi \right\} \\
&\quad\times\exp \left\{ -\frac{\mu^2}2\int_{R^2} e^{\vr(\xi)}d^2\xi\right\}. \tag{47}
\end{align*}

Note that we have further replaced the fourth-order (non-local) $2D$ gravity term by the simple local result exposed below
\begin{align*}
\, & \exp\left\{-\frac\ga2\int d^2\xi d^2\xi' e^{\vr(\xi)} e^{\vr(\xi')} \left( \overbrace{e^{-\vr(\xi)}\po^2\vr(\xi)}^{R(\xi)} \right) \times \frac{\delta^{(2)}(\xi-\xi')}{e^{\vr(\xi)}}(^{-\vr(\xi')} \po^2\vr(\xi)) \right\} \\
&= \exp\left\{-\frac\ga2\int d^2\xi d^2\xi'(\po^2\vr(\xi)) \frac{\delta^{(2)}(\xi-\xi')}{e^{\vr(\xi)}}(\po^2\vr)(\xi') \right\} \\
&\quad \sim \exp\left\{-\frac\ga2\int 
d^2\xi (\po^2\vr)^2(\xi)\right\}.\tag{48}
\end{align*}

Here we have substituted the covariant Laplace Beltrami operator Green Function by its mathematical distributional limit $\delta_{\text{cov}}^{(2)}(\xi-\xi')$, but for a back ground ``weak metric''\, $g_{ab}(\xi)=\delta_{ab}$.

In other words:
$$
\frac{\delta^{(2)}(\xi-\xi')}{e^{\vr(\xi')}} \sim \frac{\delta^{(2)}(\xi-\xi')}{1+\vr(\xi')+\dots} \sim \delta^{(2)}(\xi-\xi').
$$

It is worth to call the reader attention that we have preserved in its integrity the Liouville cosmological term $\exp\left\{-\frac{\mu^2}2\int_{R^2} e^{\vr(\xi)} d^2\xi\right\}$, since the main interest in this section is to present perturbative calculations on the $N$-point off-shell closed scattering amplitude that show us that such important term does not affects its pole singularities of those usual closed string scattering amplitudes Koba-Nielsen-Virassoro result for $\mu^2=0$; even in presence of metric fourth-order term as proposed by ours.

After collecting all the above results and remarks, we are lead to evaluate perturbativelly on the Liouville-Polyakov cosmological constant $\mu^2$, the following standard Liouville-Polyakov path integral
\begin{align*}
\tilde A(P_1,\dots,P_N) &= \int D^F[\vr(\xi)]\exp\left\{-\frac{(26-d)}{48\pi} \int_{R^2}\frac12(\po\vr)^2(\xi)d^2\xi\right\} \\
&\quad\times \exp\left\{ -\frac{\mu^2}2\int_{R^2} e^{\vr(\xi)}d^2\xi \right\} \\
&\quad\times\exp\left\{-\frac{\ga^2}2\int_{R^2}(\po^2\vr)^2(\xi) d^2\xi \right\} \\
&\quad\times \left\{ \left[ \prod_{i=1}^N e^{\vr(\xi_i)} \right] \left[\prod_{i=1}^N e^{i(P_i^\mu\cdot X_\mu(\xi_i))} \right] \right\}.\tag{49}
\end{align*}

By taking into account the momentum conservation on the scattering of the string excitations $\left(\sum\limits_{i=1}^N P_\mu^i \right)\equiv 0$, one obtains the following Liouville path integral without the $\ve$-covariant cut-off
\begin{align*}
\tilde A(P_1,\dots,P_N) &= \int_{R^{2N}} d^2\xi_1\dots d^2\xi_N \Big( \exp-\Big[ \sum_{\substack{i,j=1\\(i\ne j)}}(P_\mu^i \cdot P_\mu^j) \\
&\quad\times \left(-\frac1{4\pi}\ln|\xi_i-\xi_j|^2 \right) \Big]\Big) \\
&\quad\times\left\{ \int D^F[\vr(\xi)] \exp \left[-\frac{(26-D)}{48\pi}\int_{R^2}\frac12(\po\vr)^2(\xi) d^2\xi \right]\right. \\
&\quad \times\exp \left[-\frac{\ga^2}2\int_{R^2}(\po^2\vr)^2(\xi)d^2\xi\right] \exp\left[-\frac{\mu^2}2\int_{R^2} e^{\vr(\xi)} d^2\xi \right] \\
&\quad\times \left. \left(\exp\left[\left(\sum_{i=1}^N \vr(\xi_1)\right) -\left(\sum_{i=1}^N(P_\mu^i)^2 \frac{\vr(\xi_i)}{4\pi} \right) \right] \right) \right\}. \tag{50}
\end{align*}

It is worth to point out that the zeroth order $2D$ quantum metric $g_{ab}(\xi)=\delta_{ab}$ $(\vr(\xi)=0)$, the Liouville path integral eq.(8) produces a scattering amplitude structurally similar to the famous Veneziano-Koba-Nielsen-Virassoro closed scattering amplitude (in the Euclidean space-time $R^N$)
\begin{align*}
\tilde A_{(0)}(P_1,\dots,P_N) &= \int_{R^{2N}} d^2\xi_1\dots d^2\xi_N \left( \exp-\left[\sum_{\substack{i,j=1\\ i\ne j}}^N (P_\mu^i \cdot P_j^\mu) \cdot \left( - \frac1{4\pi}\ln |\xi_i-\xi_j|^2\right)\right] \right) \\
&= \int_{R^{2N}} d^2\xi_1\dots d^2\xi_N \left( \prod_{\substack{(i,j=1) \\ i\ne j}} \left[ |\xi_i-\xi_j|^{\frac{(P_\mu^i \cdot P_\mu^j)}{2\pi}} \right] \right).\tag{51}
\end{align*}

The key point of our study is the exactly path-integral evaluation of the fourth-order improved Liouville path integral in a perturbative expansion in the Polyakov's cosmological constant $\mu^2$. Namelly
\begin{align*}
F_{\text{Liouville}}(P_\mu^i) &\equiv \int D^F[\vr(\xi)] \exp \left[ -\frac{(26-D)}{48\pi} \int_{R^2} \frac12(\po\vr)^2(\xi)d^2\xi \right] \\
&\quad\times\exp\left[-\frac{\ga^2}2\int_{R^2} (\po^2\vr)^2(\xi)d^2\xi\right] \\
&\quad\times\left\{ \sum_{n=0}^\infty\left(-\frac{\mu^2}2\right)^n \int_{R^{2N}} e^{\vr(z_1)}\dots e^{\vr(z_N)} d^2 z_1\dots d^2 z_N \right\} \\
&\quad\times\left( \exp\left[ \sum_{i=1}^N\vr(\xi_i)(1-\frac{(P_\mu^i)}{4\pi}) \right] \right) . \tag{52}
\end{align*}

Since the above written path integral is a fourth-order Gaussian functional integral, with correlation function given explicitly by
\begin{align*}
\, & \left[ \frac{26-D}{48\pi} (-\po^2)+\ga^2(\po^2)^2\right]^{-1}(\xi,\xi')\equiv\lc^{-1}(\xi.\xi') \\
&\equiv \frac{48\pi}{26-D} \left[\left(-\frac1{2\pi}\ln|\xi-\xi'|\right) - \frac1{2\pi} \left( K_0\left( \sqrt{\frac{26-d}{48\pi\ga^2}} \, |\xi-\xi'| \right) \right) \right], \tag{53}
\end{align*}
one obtains the result below, with the path integral average Notation,
\begin{equation}
\langle\oc(\vr)\rangle_\vr=\frac1Z\left\{ \int D^F[\vr] e^{-\frac12\int\vr(\xi)\lc_\xi\vr(\xi)d^2\xi} \oc(\vr) \right\}. \tag{54}
\end{equation}

Namely: (for each perturbative order $M$ in $\mu^2$)
\begin{align*}
F_{\text{Liouville}}^{(M,N)}(P_\mu^i) &= \left\langle \prod_{j=1}^N \prod_{i=1}^M \int_{R^{2N}} < e^{\vr(z_1)} \dots e^{\vr(z_n)} \times e^{\vr(\xi_i) \left(1- \frac{(P_\mu^i)^2}{4\pi} \right)} \right\rangle_\vr \\
&= \left\{ \left( \prod_{(t,s=1)}^N \exp \left[ +\frac12\left(1-\frac{(P_\mu^t)^2}{4\pi} \right) \left(1-\frac{(P^s)^2}{4\pi} \right) \lc^{-1}(\xi_t,\xi_s) \right] \right) \right\} \\
&\quad\times\left(\prod_{(h,q)}^M\exp\left[+\frac12 \lc^{-1}(z_h,z_q) \right] \right) \\
&\quad\times \left(\prod_{t=1}^N\prod_{q=1}^M\exp\left[ +\frac12\left( 1-\frac{(P_\mu^q)^2}{4\pi} \right) \lc^{-1} (\xi_q,\xi_t) \right]\right).\tag{55}
\end{align*}

We remark now that the singularities type pole (paticle string excitation that would modify the zeroth-order Veneziano-Koba-Nielsen-Virassoro closed bosonic string dual model will come from the ``pinch''\, ultra-violet singularities $\xi_t\to \xi_s$! However from the ultra-violet behavior below depicted
\begin{align*}
\, & \lim_{\xi_t\to\xi_s'} \frac1{2\pi} K_0 \left(\sqrt{\frac{26-D}{48\pi\ga^2}} \, |\xi_t-\xi_s'| \right) \\
&\quad = \lim_{\xi_t\to\xi_s'} \left[+\frac1{2\pi}\ln |\xi_t-\xi_s'|+ \frac1{2\pi} \ln \left(\sqrt{\frac{26-D}{96\pi\ga^2}} \right) - \frac{\psi(1)}{2\pi} \right],\tag{56}
\end{align*}
one can see that there is the cancealling of the logaritmic terms on eq.(12) for the ``pinch''\, points protocol to analyze poles on string scattering amplitudes.

As a result, at each\ perturbative order $M$, these terms are expected to not contribute to the bosonic closed string spectrum eq.(10).

That is the main conclusion of our section: the Polyakov proposal of a quantum bosonic string as a theory of $D$ (ill defined!) two-dimensional massless scalar fields interacting with two-dimensional induced Liouville-Polyakov quantum gravity, (expected to be computationally effective for ``weak''\, two-dimensional quantum gravity), still does not alter the usual (tachionic) old closed bosonic string of the Veneziano-Virassoro dual model, even with a improved ultra-violet behavior as proposed by ours.

It appears thus that the introduction of induced quantum gravity in the Polyakov proposal, could not alter the usual tackionic spectrum of the bosonic string, even if one considers a somewhat ``higher order''\, Polyakov-Liouville induced $2D$ quantum gravity. As a result, one is naturally lead to consider further fermionic (intrinsic, extrinsic or supersymetric) degrees of freedom on the string world sheet ( as in refs. [1], [6]), in oder to have string's theories candidate to be physically sensible (see chapter 3).

\section{The interquark potential in Polyakov's string theory at $\frac1D$-expansion}

In this section we wishe to consider the interquark potential evaluation in Polyakov's string theory at one-loop order in a $\frac1D$ expansion in a covariant path integral framework.

Let us start our analysis by considering in a similar way to the Nambu-Goto case eq.(15-a) ([13]); the one-loop fluactuation around the string static planar reactangle configuration $X_\mu^{CL}(\xi_1,\xi_2)=\xi_1 \vec e_1+\xi_2 \vec e_2$ in the covariant Polyakov's path integral $(\Om_{(R,T)})$
\begin{align*}
\, & \frac1{2\pi \al'} \int_{\Om_{(R,T)}} d^2\xi (\sqrt g\, g^{ab} \po_a X^\mu \po_b X_\mu)(\xi) \\
&\qquad = \frac{RT}{\pi\al'} + \frac12 \int_{\Om_{(R,T)}} d^2 \xi ( \sqrt g\, g^{ab} \po_a \ov X^\mu \po_b \ov X_\mu)(\xi). \tag{57}
\end{align*}

Here the full string vector position field on $\Om_{(R,T)}$ reads as off as 
\begin{equation}
X^\mu(\xi)= X^{\mu,CL}(\xi_1,\xi_2)+ \sqrt{\pi\al'}\, \ov X^\mu(\xi_1,\xi_2). \tag{58}
\end{equation}

Note that we have used also the orthogonality condition  between the classical static quark-antiquark string field configuration and its associated closed surface transversal fluctuations $(\ov X^\mu(\xi) ; \mu=2,3,\dots,D)$
\begin{equation}
(\pi \al')^{-\frac12} \int_{\Om_{(R,T)}} d^2 \xi (\sqrt g\, g^{ab} \po_a X^{\mu,CL}, \po_b \ov X_\mu)(\xi)=0. \tag{59}
\end{equation}

After evaluating the closed field configurations covariant path integrals ([1], [9]), one obtains the result
\begin{align*}
G[C_{(R,T)}] &= \exp \left\{- \frac{RT}{\pi\al'} \right\} \\
&\quad \times \Big \{ \int D^F[\be(\xi)]\exp \Big[ - \frac{(24-D)}{48\pi} \int_{\Om_{(R,T)}} d^2 \xi \Big( \frac12 \frac{(\po\be)^2}{\be^2} \Big)(\xi) \Big] \\
&\quad \times \exp \Big[ - \frac{\mu_R^2}2 \int_{\Om_{(R,T)}} \be^2(\xi) d^2 \xi \Big]. \tag{60}
\end{align*}

In the $\frac1D$-expansion one should considers the background field decomposition around the flat metric an $\Om_{(R,T)}$ $(D\le 24)$ (Appendix C)
\begin{equation}
\be(\xi)=1+ \left( \sqrt{\frac{48\pi}{24-D}} \right) h(\xi) \tag{61}
\end{equation}
which leads to the one-loop result
\begin{align*}
G[C_{(R,T)}] &= \exp \left\{ -RT \left( \frac1{\pi\al'} + \frac{\mu_R^2}2 \right) \right\} \\
&\quad \times \det_{\Om_{(R,T)}}^{-\frac12} \left[ -\po^2 + \frac{48\pi}{24-D} \, \mu_R^2 \right] \tag{62}
\end{align*}
where we have supposed the ``tad-pole''\, dimensional regularization zero condition imposed on ours evaluation
\begin{equation}
\int_{\Om_{(R,T)}} d^2\xi h(\xi)=0. \tag{63}
\end{equation}

On basis of the above result and eq.(35) with the identification $\frac1{\ga^2} = \frac{48\pi}{24-D}\, \mu_R^2$ on that equation eq.(35) one obtains the large $R$ limit of string vacuum energy (the interquarkl potential in this model)
\begin{align*}
\overset{\text{Polyakov}}{V(R)} &= R\left( \frac1{\pi\al'}+\frac{\mu_R^2}2 \right) \\
&\quad + \Big\{ - \frac\pi{6R}+ \frac1{2\mu_R^2(\ec)} \\
&\quad - \frac1{4\pi(\mu_R(\ec))^4} (R \ln(4\pi e^{-\hat\ga})) \\
&\quad + \frac1{4\pi} \cdot \frac1{\mu_R^4(t)} (R\ln R) \\
&\quad + \Big( \frac1{16\pi^{s/2} \mu_R^2(\ec)} \Big) R \Big\}. \tag{64}
\end{align*}

It is worth to compare the very similar functionals structure of the interquark potential on the Nambo-Goto string and on the Polyakov's string at $D\to-\infty$.

Another important remark to be called the reader attention is that one which gives an representation for the scalar meson propagator entirelly in terms of a loop path integral, added with the Nambo-Goto and Polyakov's correction defined by the minimal surface bounded by the above loop (an euclidean space-time trajectory of a pair quark-antiquark, at least for $N_c\to\infty$) in this string model for the $QCD(SU(\infty))$ vacuum (see next section)
\begin{align*}
\, & F_{\ec_{QCD}}(|x-y|) \sim \langle(q\bar q)(x) (q\bar q)(y)\rangle_{QCD(SU(\infty))} \\
&= \int_{\ec_{QCD}}^\infty \frac{dt}t \Big\{ \int_{\re^D} \frac{d^Dk_1 d^Dk_2}{(2\pi)^D} e^{-ik_1x} e^{-ik_2y} \int_{\re^D} dz \int_0^t d\sigma_1 \int_0^t d\sigma_2 \\
&\quad \times \Big[ \int_{C^\mu(0)=C^\mu(t)=z^\mu} D^F[C^\mu(\sigma)] e^{+ik_1C(\sigma_1)}  e^{+ik_2C(\sigma_2)} \exp\left(-\frac12 \int_0^t (C^\mu(\sigma))^2 d\sigma\right) \\
&\quad \times \int_{0^+}^\infty dA \Big( \exp \Big\{ - \frac{(26-D)}{48\pi} \Big[ \int_0^A d\xi_1 \int_0^t d\xi_2 (\frac12 (\po \bar\vr)^2+ \frac{\mu^2}2 e^{\bar\vr}) (\xi_1,\xi_2) \Big] \Big\} \\
&\quad \times \exp \Big \{ - \frac1{2\pi\al'} \int_0^A d\xi_1 \int_0^t d\xi_2 (e^{\bar\vr})(\xi_1,\xi_2) \Big\} \Big) \Big] \Big\}. \tag{65}
\end{align*}

Here
\begin{equation}
\bar\vr(\xi) = \ln [h(X^{\mu,CL}(\xi))] \tag{66}
\end{equation}
and the minimal surface is an the conformal gauge (see discussion on \S 2).

We will attempt to evaluate eq.(65) elsewhere.

Finally eq.(65) preserves its structural functional form for strings moving on curved space-time with metric $G_{\mu\nu}(X^\al)$ ([1]). We only need to make the replacements:
\begin{equation}
h_{ab}^{CL}(X^{\mu,CL}(\xi))=\po_a X^{\mu,CL}(\xi) G_{\mu\nu}(X^{CL}(\xi)) \po_b X^{\nu,CL}(\xi) \tag{67-a}
\end{equation}
and the minimal surface equation on the ambient Riemann Manifold space-time
\begin{align*}
\Delta_h X^{\mu,CL} &+ h^{ab}(X^{\mu,CL}(\xi)) \Ga_{\be\ga}^\mu(X^{\mu,CL}(\xi)) \\
&\quad \times \po_a X^{\be,CL}(\xi) \po_b X^{\ga,CL}(\xi)=0 \tag{67-b}
\end{align*}
\begin{equation}
X^{\mu,CL}(\xi) \Big\vert_{\po\vr}=C^\mu(\xi), \tag{67-c}
\end{equation}
here $\Ga_{\be\ga}^\mu$ are the Christoffel symbols associated to the ambient Manifold $G_{\mu\nu}(X^\al)$.

Let us conjecture that even in the presence of a Blach-Hole metric, One still has color charge confinement and string structures for $QCD(SU(\infty)!)$.

Anyway we study the above mentioned case of an ambient external Riemannian metric at one-loop case but now on the more easily handled Nambo-Goto string path integral.

In this case we have the one-loop approximations (weak fluctuations of the space-time metric around the flat metric)
\begin{equation}
X^\mu(\xi_1,\xi_2)=\xi_1\vec e_1+\xi_2\vec e_2+(\sqrt{\pi\al'}) Y^\mu(\xi_1,\xi_2) \tag{68-a}
\end{equation}
\begin{equation}
G_{\mu\nu}(X^\al(\xi_1,\xi_2) \approx \delta_{\mu\nu} -
\frac{\pi\al'}3 [\La(\delta_{\al\be} \delta_{\mu\nu}-\delta_{\mu\be}\delta_{\al\nu})](Y^\al Y^\be)(\xi). \tag{68-b}
\end{equation}

Here the four index Riemann Tensor is taken to be of an extended deSitter-Metric with negative cosmological constant $\La$.

In this case, straightforward evaluations lead us to the $2D$-intrinsic metric components results:
\begin{align*}
h_{00}(\xi_1,\xi_2) &= \Big[ (1+\po_{\xi_1} Y^\mu \po_{\xi_1} Y_\mu)- \frac{\pi\al'}3 \La(Y^\mu Y_\mu) \Big](\xi_1,\xi_2) \tag{69-a}
\\ 
h_{11}(\xi_1,\xi_2) &= \Big[ (1+\po_{\xi_2} Y^\mu \po_{\xi_2} Y_\mu)- \frac{\pi\al'}3 \La(Y^\mu Y_\mu) \Big](\xi_1,\xi_2) \tag{69-b} \\
h_{01}(\xi_1,\xi_2) &= h_{10}(\xi_1,\xi_2)=\po_{\xi_1} Y^\mu \po_{\xi_2} Y_\mu \tag{69-c}
\end{align*}

By keeping only quadratic terms on $\al'$, one thus gets the one-loop action:
\begin{align*}
S & \sim \frac1{2\pi\al'} \Big\{ \int_{\Om_{(R,T)}} d^2\xi \Big[ 1+(\pi\al')(\po Y^\mu)^2-\frac{2\pi^2}3(\al')^2\La(Y^\mu)^2 \Big]\Big\}(\xi_1,\xi_2) \\
&= \frac1{2\pi\al'} RT+\frac12 \Big\{ \int_0^R d\xi_1 \int_0^T d\xi_2 \Big[(\po Y^\mu)^2+M^2(Y^\mu)^2 \Big]\Big\}(\xi_1,\xi_2). \tag{70}
\end{align*}

Here the effective mass parameter $M^2$ is given explicitly
\begin{equation}
M^2=-\frac{4\pi^2}3(\al')\La>0.\tag{71}
\end{equation}

As a result, one gets the following expression for the interquark potential in this one loop evaluation
\begin{align*}
V(R)= \frac R{2\pi\al'} &+ \Big\{ \Big(-\frac{(D-2)}6 \cdot \frac\pi R \Big)+ \frac{(D-2)}2 M_{\text{ren}}^2(\ec) \\
&- \left( \frac{(D-2)}{4\pi}(M_{\text{ren}}^2(\ec))^2\right) R\ln(4\pi e^{-\hat\ga}) \\
&+ \frac{(D-2)}{4\pi} (M_{\text{ren}}^2(\ec))^2(R\ln R) \\
&+ \left( \frac{(D-2)}{16\pi^{s/2}} \right) R \Big\}. \tag{72}
\end{align*}

We point out again the comnonality of the functional terms of eq.(72), eq.(64) and eq.(41).

The result eq.(72) may be the reason behind of the conjecture about strings moving in background Riemman geometries should represents some regime of the $QCD(SU(\infty))$ string. However more extensive work on this equivalence is needed.

\section{Area functional as vacuum structures for $QCD(SU(\infty))$}

One of the most interesting speculations on $QCD(SU(\infty))$ is that the vacuum functional loop wave functions are area functionals, which by its turn signals out the color charge confinement of quarks and gluous ([1], [2]).

The first step to arrive at such results, one should consider the loop wave equation for $QCD(SU(\infty))$ for frozen quark spin dynamics, which means the Migdal-Makeenko loop wave equation for the $QCD(SU(\infty))$ Wilson Loop for non self-intersecting loops ([2]).
\begin{equation}
\po_\mu^z \frac\delta{\delta\sigma_{\mu\nu(z)}}(W[C_\mu])=(g_\infty^2(a^2))\oint_C \delta^{(D)}(z_\mu-C_\mu(\sigma))W[C_\mu]. \tag{73}
\end{equation}
Here $g_\infty^2(a^2)$ should be some sort of large $N_c\to\infty$ dimensionally transmuted $QCD$ coupling constant.

Let us thus search for a solution of eq.(68) by means of a loop distributional calculus
\begin{equation}
W[C:= \exp \left\{ -b \int_S d\sigma^{\al\be}(x) \delta^{(D)}(x-y) d\sigma_{\al\be}(b)\right\}. \tag{74}
\end{equation}

Let us evaluate te parameter $b$, in order tod eq.(74) satisfies eq.(73).

Firstly we note that
\begin{equation}
\frac\delta{\delta\sigma_{\mu\nu}(z)} W[C]=W[C] \left\{-2b\int_S \delta^{(D)} (z-y)d\sigma^{\mu\nu}(y)\right\}.\tag{75-a}
\end{equation}

We also have the thistributional result at the surface boundary $\po S=C$
\begin{equation}
\po_\mu^z \delta^{(D)} (z_\mu-X_\mu(\xi))=-\frac\delta{\delta X_\mu(\xi)} \delta^{(D)}(z_\mu-X_\mu(\xi)).\tag{75-b}
\end{equation}
Note that
\begin{equation}
d\sigma^{\al\be}(x)= \zeta^{\al\be}(X(\xi)) \delta^{(D)}(x-X(\xi))d^2\xi \tag{75-c}
\end{equation}
which produces the result $(S=X_\mu(D))$
\begin{equation}
\po_\mu^z \frac\delta{\delta\sigma_{\mu\nu}}(\xi) W[C] =+2b W[C] \left[ \int_D\delta^{(D)}(z_\mu-y_\mu(\xi)) \left( \frac\delta{\delta X_\mu(\xi)} \zeta^{\mu\nu}(X(\xi)) \right) d^2\xi \right]. \tag{76}
\end{equation}

Note that the loop functional derivation of the surface area tensor $\zeta^{\mu\nu}(X(\xi))$ is made at the boundary of $S$, which is $C$ (a non self-intersecting loop!). Its result reads ([1])
\begin{equation}
\zeta^{\mu\nu}(X(\xi))\Big|_{X_\mu(\xi)=C_\mu(\xi)} = (C^\mu \dot C^\nu - \dot C^\mu C^\nu)(\sigma). \tag{77}
\end{equation}
It yield thus
\begin{equation}
\frac\delta{\delta C_\mu(\sigma)}([C^\mu \dot C^\nu -\dot C^\mu C^\nu](\sigma)) = 2\delta^{(\ve)}(0) \dot C^\nu(\sigma). \tag{78}
\end{equation}
Here $\delta^{(\ve)}(0)$ is a regularized form of the delta function at zero argument ([1]), the so called Nielsen-Olesen $QCD(SU(\infty))$ vacuum.

Collecting togheter all the below written distributional loop-surface operational calculus, one gets finalls
\begin{equation}
\po_\mu^z \frac\delta{\delta\sigma_{\mu\nu}(\xi)} W[C_{zz}]=W[C_{zz}](4b\delta^{(\ve)}(0)) \left( \oint_{C_{zz}}\delta^{(D)}(z_\mu-C_\mu(\sigma)) \dot C^\nu(\sigma)d\sigma \right). \tag{79}
\end{equation}

Let us thus choose the $b$ parameter on eq. (74) (in the notation of ref. ([1])) by introducing a dimensional $QCD$ coupling constant
\begin{equation}
b= \frac{\lim\limits_{N_c\to\infty}(g^2 N_c)}{4\delta^{(\ve)}(0)} := \frac{g_\infty^2(a)}{a^2}. \tag{80}
\end{equation}

By using again the operational distributional surface rule for non self-intersecting surfaces
\begin{equation}
\delta^{(D)} (X_\mu(\xi)-X_\mu(\xi')):= \delta^{(2)}(\xi-\xi')\Big/ (h(X(\xi))^{D/8} \tag{81}
\end{equation}
and the relationship
$$
\zeta^{\mu\nu}(X^\al(\xi)) \cdot \zeta_{\mu\nu}(X^\al(\xi))=1
$$
\begin{equation}
d\sigma^{\mu\nu}(X^\al(\xi))=\sqrt{h(X^\al(\xi))} \cdot \zeta^{\mu\nu}(X^\al(\xi)) \tag{82}
\end{equation}
one gets the Nambo-Goto area functional as formal and non perturbative wave functionals for the $QCD(SU(\infty))$ vacuum at $D=4$
\begin{align*}
W[C] &= \exp \Big\{ - \frac{g^2(a)}{a^2} \int_D d^2 \xi' \int_D d^2\xi (\sqrt{h(X(\xi))}) \cdot \zeta^{\mu\nu}(X(\xi)) \\
&\quad \times \frac{d^{(2)}(\xi-\xi')}{(\sqrt{h(X(\xi))})} \, (\sqrt{h(X(\xi'))}) \cdot \zeta_{\mu\nu}(X(\xi')) \Big\} \\
&= \exp \left\{-\frac{g^2(a)}{a^2} \int_D d^2\xi \sqrt{h(X(\xi))} \right\}. \tag{83}
\end{align*}

The identification with string theory wave functionals should be made through the relationship
\begin{equation}
\frac{g_\infty^2(a^2)}{a^2} = \frac1{2\pi\al'}.\tag{84}
\end{equation}

\newpage

\noindent{\Large \bf Appendix A: The distributional limit of the Epstein \linebreak function}

\vskip .2in

Let us try to evaluate (in some yet undiscovered asymptotic distributional theory) the limit of large a of the so called Epstein function for $s\in R$
\begin{align*}
S_{\text{epstein}}(s,a^2) &= \sum_{n=1}^\infty \frac1{(n^2+a^2)^s} \\
&= \frac1{\Ga(s)} \left( \int_0^\infty U^{s-1} e^{-a^2U^2} e^{-Un^2} dU \right) \\
&= \frac1{\Ga(s)} \left( \int_0^\infty U^{s-1}\cdot e^{-a^2U^2} \cdot 
Tr_{\left\{ \begin{matrix} 
C^2([0,1]) \\ \text{Dirichlet} \end{matrix} \right\} }
\left(e^{-U\left(\frac{-d^2}{d \xi'} \right)} \right) \right).\tag{A-1}
\end{align*}

For $a\to\infty$,k certainly $U\to0$ on the $U$-integrand, and the Seeley asymptotic expansion for the second-order operator $-\frac{d^2}{dz^2}$ holds true.

Theta is:
\begin{align*}
S_{eps}(s,a^2) & \overset{a^2\to\infty}{\sim} \, \frac1{\Ga(s)} \left[ \int_0^\infty dU\cdot U^{s-1} e^{-Ua^2}\, \frac1{4\pi U^{1/2}} \right] \\
& \overset{a^2\to\infty}{\sim} \frac1{4\pi \Ga(s)} \left[\int_0^\infty dU\cdot U^{s-\frac32} e^{-Ua^2} \right] \\
& \overset{a^2\to\infty}{\sim} \frac1{4\pi \Ga(s)} \left[ \frac{\Ga(s-\frac32+1)}{(a^2)^{s-\frac32+1}} \right]
\end{align*}

\newpage

\noindent{\Large \bf Appendix B: Integral Evaluation}

\vskip .2in

Let us elementarly evaluate the following integral
\begin{align*}
\, & \int_0^1 dx(1-x) \frac1{(a^2+xb)^{3/2}}=\int_0^1 \frac{dx}{(a^2+xb)^{3/2}}-\int_0^1 dx \frac x{(a^2+xb)^{3/2}} \\
& \overset{(x=\frac{v-a^2}b)}{=} \frac1b \left( \int_{a^2}^{a^2+b} dv\cdot v^{-\frac32} \right)-\frac1{b^2}  \left( \int_{a^2}^{a^2+b} dv\cdot v^{-\frac12} \right) \\
&\quad - \frac{a^2}{b^2} \left( \int_{a^2}^{a^2+b} dv\cdot v^{-\frac32} \right) \\
&= \left[(a^2+b^2)^{-\frac12}\left(-\frac2b+\frac{2a^2}{b^2}\right)\right] \\
&\quad + \left[\frac2{ab}+\frac{2a}{b^2}-\frac{2a}{b^2}\right]-\frac2{b^2}(a^2+b)^{1/2} \\
&= \left(-\frac{2b+2a^2}{b^2} \right) \frac1{(a^2+b)^{1/2}}+\frac2{ab} - \frac2{b^2}(a^2_+b)^{1/2}.\tag{B-1} \\
\end{align*}

As a consequence $(\mu^2=n^2)$
\begin{align*}
\int_0^1 dx(1-x)\frac1{(\mu^2+xa^2)^{3/2}} &= -\left(\frac2{a^2}\right) \left(\frac1{(\mu^2+a^2)^{1/2}} \right) \\
&\quad+\frac1{a^4} \cdot \frac{2\mu^2}{(\mu^2+a^2)^{1/2}} \\
&\quad+ \frac2{a^2 \mu}-\frac2{a^4} (\mu^2+a^2)^{1/2}. \tag{B-2}
\end{align*}

\newpage

\noindent{\Large \bf Appendix C: On the perturbative evaluation of the bosonic string closed scattering amplitude on Polyakov's framework}

\vskip .2in

The fundamental observable on Polyakov's bosonic string theory ([1], [2]) is the closed scattering amplitude which is given explicitly by the covariants $2D$ induced quantum gravity path integral in moments space
\begin{align*}
\, &\tilde A(P_\mu^1,\dots,P_\mu^N)=\prod_{j=1}^N \left[\int d^2\xi_j \times \left(\frac1Z\int D^F[\bar\rho(\xi)] \right. \right. \\
&\times \exp \left\{ - \frac{(26-D)}{12\pi} \int_{\re^2} \left(\frac12\left(\frac{\po_a\bar\rho}{\bar\rho} \right)^2\right)(\xi) d^2\xi - \frac12\mu_R^2 \int_{\re^2} \bar\rho^2(\xi)d^2\xi \right\} \\
&\times \exp \left\{ \sum_{j=1}^N 2\ln(\bar\rho(\xi_j))\right\} \\
&\times \left. \left. \exp \left\{-\sum_{(i,j)=1}^N(P_\mu^i \cdot P_\mu^j)_{R^N} K^{(\ve)}(\xi_i,\xi_j) \right\} \right) \right]. \tag{C-1}
\end{align*}

Here the covariant regularized Polyakov's Green function is given explicitly by eq.(3) (with the identification $(\bar\rho(\xi))^2=e^{\vr(\xi)}$).

After substituting eq.(3) into eq.(1-a), one obtain the outcome expressed now as a sigma model like perturbativelly renormalizable path integral defined by a trully Feynman measure $D^F[\bar\rho(\xi)]$ ([1], [9]).

We have thus to evaluate in term of the $\frac1D$-expansion, the natural perturbative parameter expansion on two-dimensional quantum gravity applied for quantum strings with domain parameter being $\re^2$ (closed string) (see footnote)
\begin{equation}
\bar\rho(\xi)=1+\left(\frac{12\pi}{26-D} \right)^{1/2} h(\xi). \tag{C-2}
\end{equation}

Here we have considered our classical background metric, the flat metric $g_{ab}(\xi)=1\delta_{ab}$ on $\re^2$, as it should be in Einstein like gravitation theories.

At one-loop order, one arrives at the following path integral (with the normalization factor $Z=1$).

The full $\frac1D$ expanded path integral is given by
\begin{align*}
Z &= \int D^F[h]\exp\left\{-\frac12\int_{\re^2} \left[(\po h)^2\left(\sum_{n=1}^\infty(-1)^{n-1} n \ve^{n-1} h^{n-1}\right) \right] (\xi) d^2\xi \right\} \\
&\qquad\times \exp \left\{-\mu^2\int_{\re^2} h^2(\xi)d^2\xi\right\}\tag{C-3}
\end{align*}
with the orthogonal constraint of the fluctuating piece $h(\xi)$ in relation to the flat $\re^2$ background
\begin{equation}
\int_{\re^2} h(\xi)\cdot 1\cdot d^2\xi=0. \tag{C-4}
\end{equation}

From the usual renormalization power counting interactions of the form $((\po h)^2 h^m)(\xi)$ have the renormalization index $r=\left(\frac{(D-2)}2\right) b+\frac{(D-1)}2 f+\delta-D$ for an interaction of general for $g(\po)^\delta \phi^b \psi^F$. In our case $g=\ve$, $D=2$, $f=0$, $\delta=2$, $b=m$, which leads to the theory's renormalizability $r=0$ in $\re^2$.
\begin{align*}
A(P_\mu^1,\dots,P_\mu^N) &= \frac12\left\{ \int D^F[h] \exp \left(
-\frac12\int_{R^2}[(\po h)^2+\mu^2 h^2](\xi)d^2\xi \right) \right\} \\
&\quad\times \left\{ (\ve)^{(\sum\limits_{j=1}^N(P_\mu^j)^2)} \times \int_{\re^2} \prod_{j=1}^N d^2 \xi_j \left( \prod_{1<j} |\xi_i-\xi_j|^{\frac{(P_\mu^i\cdot P_\mu^j)}{2\pi} \re^N} \right) \right\} \\
&\quad \times \left[ \prod_{j=1}^N \exp \left( 2 \left( \frac{12\pi}{26-D} \right)^{1/2} \left( 1- \frac{(P_\mu^j)^2}{4\pi} \right) h(\xi_j) \right) \right].\tag{C-5}
\end{align*}

Here we remark the use of the one-loop approximation on the object $\left( \ve=\left(\frac{12\pi}{26-D}\right)^{1/2} \right)$
\begin{align*}
\, & \left[ \prod_{j=1}^N (\bar\rho(\xi_j))^{2(1-\frac{(P_\mu^j)^2}{4\pi})} \right]
 = \left[ \prod_{j=1}^N e^{2(1-\frac{(P_\mu^j)^2}{4\pi})\ln(1+\ve h(\xi_j))} \right] \\
&\quad \overset{\text{one-loop order}}{\sim} \left\{ \prod_{j=1}^N \exp \left[ 2\ve \left(1- \frac{(P_\mu^j)^2}{4\pi} \right) h(\xi_j) \right] \right\}. \tag{C-6}
\end{align*}

At this point one can see that at a perturbativelly renormalized ``$\frac1D$ expansion''\, $(D\to-\infty$) ([1], [9]), the contribution of the Liouville-Polyakov"'s dynamics is a multiplication factor as given below ( $\sum\limits_{j=1}^N(P_\mu^j)^2=0$ for a physical elastic string excitation scattering)
\begin{align*}
\tilde A(P_\mu^1,\dots,P_\mu^N) &= \int_{\re^2} d^2\xi_j\left[ \prod_{i<j}^N |\xi_i-\xi_j|^{(\frac{P_\mu^i\cdot P_\mu^j}{2\pi})_{\re^N}} \right] \\
&\quad \exp \left\{ \sum_{i<j=1}^N \frac{\overbrace{24\pi}^{2\ve^2}}{(26-D)} (1-(P_\mu^i)^2)(1-(P_\mu^j)^2) \right.\\
&\quad \left. \times \left( \frac1{2\pi} K_0 (\mu_R |\xi_i-\xi_j|)\right) \right\}. \tag{C-7}
\end{align*}

By considering $\ve\to0$ $(D\to-\infty)$, one re-obtains the well-known A.M. Virassoro-B. Sakita closed scattering from the old dual models.

Another useful remark is that the behavior of eq.(5-a) for $\xi_i\to\xi_j$ is the same of the Virassoro-Sakita result, implying, thus, that by just taking into account the Liouville-Polyakov degree of freedom on Bosonic Srings does not alter the poles of the scattering amplitudes, so leasing to the same drawbacks of the old theory.

We conclude that nonperturbative effects (or others intrinsic degrees of freedom ([1])) must be taken into account to remove the tachion from the Polyakov Bosonic string spectrum ([1]).

\newpage

\centerline{\Large\bf Note Added}

\vskip .2in

We wishe to make some remarks to clarify the text:

\vskip .1in

\noindent
1) It is important to remark for the reader to not confuse our perturbative expansion on terms of the string constant tension $\al'$ with the string unitarity corrections on the underlying $S$-matrix for the string's excitations which is taken into account by considering the topological-genus expansion of M. Virassovo--A. Alexandrini in the string path integral.

\vskip .1in

\noindent
2) It is also charifying to remark that the equivalence between the extrinsic string coupling constant in our paper is made on basis of ref. Luiz C.L. Botelho -- ``String wave equations in Polyakov's path integral framework'' -- J. Math Phys. Vol 30, No. 9, September 1989, eqs. (25b), (29-a), (29-b) Appendix B; and the study of ref. A.I. Karanikos and C.N. Ktorides -- ``Surface Self-Intersections and Instantons'' -- Physics Letters 235B, No. 1,2, 1990 -- eqs. (17), (29), where this relationship is taken on a regularized form for the surface's self-intersections. See also discussions on Section 6 of this paper.

\vskip .1in

\noindent
3) It is worth to call attention that scalar amplitudes on string theory are not in principle tachians amplitudes. This kind of identification is done, on the Polyakov's path integral framework, only after the writing of the associated poles on-shell of the strings scalar amplitudes. See for instance the basic reference:

Yadin Goldschimidt and Chong-I Tan:

``Saddle point analysis of Pomeron Singularities in Polyakov's string theory'' -- Phys. Rev. Lett 114B, No. 4, 1982, eq.(25).

\vskip .1in

\noindent
4) The well-known statistical mechanics of Ising Lattice Models at $\frac1D$ expansion is generalized easily to A.M. Polyakov's string path integral, after the extrinsic-ambient string degrees of freedom are functionally integrated out. See Luiz C.L. Botelho -- ``A Scattering Amplitude in Quantum Geometry of fermionic strings'',  Physics Letters 152B, No. 5,6, eq.(16), (17) and B. Durhuus, P. Olesen, and J.L. Petersen, Nucl. Phys. B198, 159, (1982).

\vskip .1in

\noindent
5) On the possible (but not mysterious!) connection between the target space dimension and the rank of the QCD gauge group on the Polyakov's string path integral see ref. Luiz C.L. Botelho -- ``The $D\to-\infty$ Saddle-point spectrum analysis of the open bosonic Polyakov string in $R^D\times SO(N)$, Physical Review 35D, No. 4, (1987), pp. 1515--1518, and cited references.

\vskip .3in

\noindent
{\bf Acknowledgments:}
The author is thankfull to Professor Waldyr Rodriguez-IMEC/UNICAMP for being sponsor of an author research visit at IMEC-UNICAMP under financial fellowship of CNPq. Senior Pos Doctoral Fellowship

\vskip .5in

\noindent
{\bf REFERENCES}

\begin{itemize}

\itemsep= -2pt

\item[{[1]}] Luiz C.L. Botelho, Methods of Bosonic and Fermionic Path Integrals, Nova Science Publisher, Inc., 2009.

\item[{[2]}] A.M. Polyakov, Gauge Fields and Strings, Harwoud Academic Publisher, Chur., (1987). 

\item[{[3]}] Michio Kako, Introduction to Superstrings, Springer Verlag, 1988.

\item[{[4]}] Luiz C.L. Botelho, Lecture Notes in Applied Differential Equations of Mathematical Physics, World Scientific, 2008.

\item[{[5]}] M.J. Duff, Partial Differential Equations, Toronto Press, (1956).

\item[{[6]}] M.B. Green, J.H. Schwarz \& E. Witten, Superstring theory, vol. 2, Cambridge Monographs on Mathematical Physics, (1996).

\item[{[7]}] C. Itzykson \& J.M. Drouffe, Statistical Field theory, vol. 1, Cambridge Monographs on Mathematical Physics, (1991).

\item[{[8]}] Luiz C.L. Botelho, Mod. Physics Letters A, vol. 20, No. 12, (2005).

\item[{[9]}] Luiz C.L. Botelho, ISRN High Energy Physics, vol. 2012, Article ID 674985, doc. 10.5402/2012/674985. (Research Article) -- chapter 1.

\item[{[10]}] J. Teschner, Class. Quant. Grav., 18, (2001), R153--R222.

\item[{[11]}] Adel Bilal, F. Ferrari, S. Klevitson, arxi No. 1310.1951v2, [hep-th], (1800--2013).

\item[{[12]}] Luiz C.L. Botelho, Gen Relativ Gravit (research article), DOI 10.1007/S, 10714-012-1372-1, (May/11/2012).

\end{itemize}

\end{document}